\newcommand{\CLASSINPUTinnersidemargin}{18mm}
\newcommand{\CLASSINPUToutersidemargin}{12mm}
\newcommand{\CLASSINPUTtoptextmargin}{20mm}
\newcommand{\CLASSINPUTbottomtextmargin}{25mm}
\documentclass[10pt,conference, twoside,a4paper]{IEEEtran}

\usepackage[utf8]{inputenc}

\usepackage{times}

\renewcommand{\abstractname}{Abstract}    

\usepackage{graphicx}
\usepackage{subfigure}
\DeclareGraphicsExtensions{.png,.eps,.ps,.pdf}

\usepackage{fancyhdr}
\usepackage{kantlipsum}

\usepackage{url}

\usepackage[left=2cm,top=2.5cm,right=2cm, bottom=2.5cm]{geometry}

\usepackage[usenames]{color}

\usepackage{eso-pic}

\begin{document}

\AddToShipoutPictureBG*{%
  \AtPageUpperLeft{%
    \setlength\unitlength{1in}%
 	\hspace{2cm}
 	 	\makebox(0,-2)[l]{
			\begin{tabular}{l r} 
			\multicolumn{1}{p{12cm}}{\vspace{-0.3cm}\includegraphics[scale=0.50]{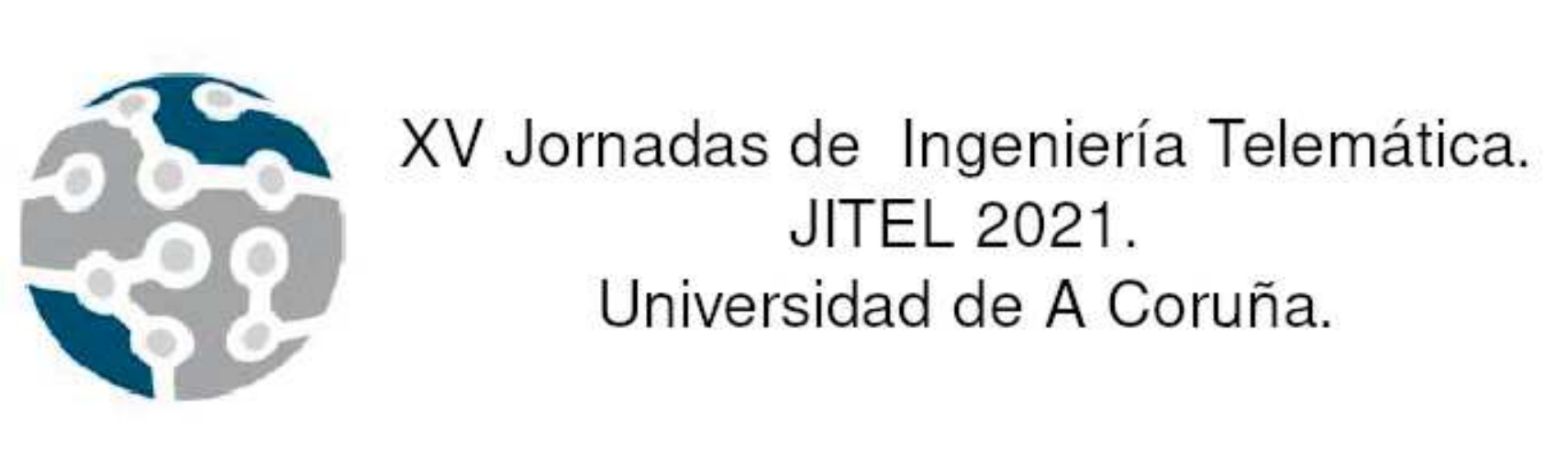}} & 						\multicolumn{1}{p{4cm}}{\raggedleft\small\usefont{T1}{phv}{m}{it} Actas de las XV Jornadas de Ingeniería Telemática\\ (JITEL 2021),\\ A Coruña (España), \\27-29 de octubre de 2021. \vspace{0.2cm} \\ISBN} \tabularnewline 
			\end{tabular}
 }
}}

\AddToShipoutPictureBG*{%
  \AtPageLowerLeft{%
    \setlength\unitlength{1in}%
    \hspace*{\dimexpr0.5\paperwidth\relax}
    \makebox(0,1.3)[c]{\footnotesize\usefont{T1}{phv}{m}{} This work is licensed under a \underline{\textcolor{blue}{Creative Commons 4.0 International License}} (CC BY-NC-ND 4.0)}%
}}

\title{\vspace{2cm}Towards integrating hardware Data Plane acceleration in Network Functions Virtualization}

\author{\IEEEauthorblockN{David Franco$^{1}$, Asier Atutxa$^{1}$, Jorge Sasiain$^{1}$, Eder Ollora$^{2}$, Marivi Higuero$^{1}$, Jasone Astorga$^{1}$, Eduardo Jacob$^{1}$}
\IEEEauthorblockA{$^{1}$Department of Communications Engineering, University of the Basque Country (UPV/EHU). 48013 Bilbao, Spain.\\
\{david.franco, asier.atutxa, jorge.sasiain, marivi.higuero, jasone.astorga, eduardo.jacob\}@ehu.eus}
\IEEEauthorblockA{$^{2}$DTU Fotonik, Technical University of Denmark. Kongens Lyngby, Denmark.\\
eoza@fotonik.dtu.dk}
}

\maketitle

\begin{abstract}
\textbf{This paper proposes a framework for integrating data plane (DP) acceleration within the Network Functions Virtualization (NFV) architecture. Data plane programming (DPP) proves to be beneficial for NFV environments, as it provides full packet forwarding flexibility through the use of self-designed algorithms. Additionally, DPP provides high-performance networking, as the DP can be configured to execute specific functions on dedicated hardware. We present an integration of the DP acceleration within the ETSI NFV architecture that leverages custom DP functions implemented in hardware switches using P4 language. Besides, OpenStack and Kubernetes are used as Virtualized Infrastructure Managers (VIMs) and Open Source MANO (OSM) as the Management and Orchestration (MANO) element.
}
\end{abstract}

\section{\uppercase{Introduction}}
In the last years, Network Functions Virtualization (NFV) and Software-Defined Networking (SDN) have changed the framework for the deployment of services. On the one hand, NFV allows Network Functions (NFs) to be deployed as Virtual Network Functions (VNFs) over a commercial off-the-shelf hardware infrastructure. On the other hand, SDN solves the problem of having a vendor-specific control plane (CP) in network devices, allowing the definition of custom CPs designed by the network operator.
However, full packet forwarding flexibility is given by data plane programming (DPP), which can forward packets following self-designed algorithms and perform custom actions to packets with user-defined formats. Programming Protocol-Independent Packet Processors (P4) is a DPP language that provides a high abstraction level to define packet processing pipelines.
DPP also provides high performance networking, as the data plane of network devices can be configured to execute specific functions on dedicated hardware. The addition of DPP to NFV through hardware network devices enhances the performance in the communication among VNFs thanks to line-rate packet processing capability and the ability to offload certain compute-intensive functions.

This paper proposes an integration of the DP acceleration within the standard ETSI NFV architecture, using OpenStack and Kubernetes to manage the NFV Infrastructure (NFVI), and Open Source MANO (OSM) as the top-level NFV orchestrator. The framework integrates the lifecycle of P4-enabled hardware switches in the NFV architecture to enable the automatic offloading of NFs to the DP.

\section{\uppercase{Related work}}





Different works have been presented in the literature regarding the data plane (DP) function offloading in NFV. For instance, \cite{piaffe,kathara} introduce two frameworks to offload VNF processing to P4-based Network Interface Cards (NICs) and software switches respectively. The former decomposes VNFs into small embedded NFs that are offloaded to the DP, and the latter runs NFs as P4 programs on Docker containers that implement P4 targets.

Authors in \cite{p4nfv} propose an ETSI NFV-based architecture that allows the instantiation of P4-based NFs. They explain the necessity of reconfiguring the DP when multiple users are sharing the same physical P4 target and they test their architecture over P4 software switches. 
In this sense, \cite{mtpsa,microp4,p4click} provide different approaches to achieve DP modularity and allow a transparent reconfiguration of the DP. The majority of them require modifications of the P4 compiler, or even in the architecture of the P4 target. For instance, \cite{microp4,p4click} present abstraction layers and mechanisms to create modular P4 programs that can then be merged. 

Authors in \cite{dppx,offloading} describe how to integrate P4 within OpenStack, by modifying the Neutron module to allow the offloading of some VNF functions to the DP of P4 software switches. They classify their VNF offloading according to the acceleration techniques defined by the ETSI NFV \cite{etsiaccel}. However, none of these approaches considers the use of hardware-based targets.

This paper focuses on the advantages of integrating P4-enabled hardware switches in NFV to automate the offloading and acceleration of custom NFs to the DP.

\section{\uppercase{Proposed framework for DP acceleration in NFV}}
This section describes the proposed integration of the DP acceleration within the ETSI NFV architecture. The proposed framework allows the users to create network services (NSs) that leverage custom DP functions implemented in P4 language. For this purpose, physical P4-enabled switches are integrated into the NFVI.


\subsection{Functional description}
In this subsection we present an overview of the proposed technologies and approaches to enable the deployment of NSs combining traditional NFs ---VNFs and Container Network Functions (CNFs)--- with our novel proposed concept of P4 Network Functions (P4NFs). We propose the deployment of such P4NFs over P4 switches.

\begin{figure}[tb]
\centerline{
\includegraphics[width=\linewidth]{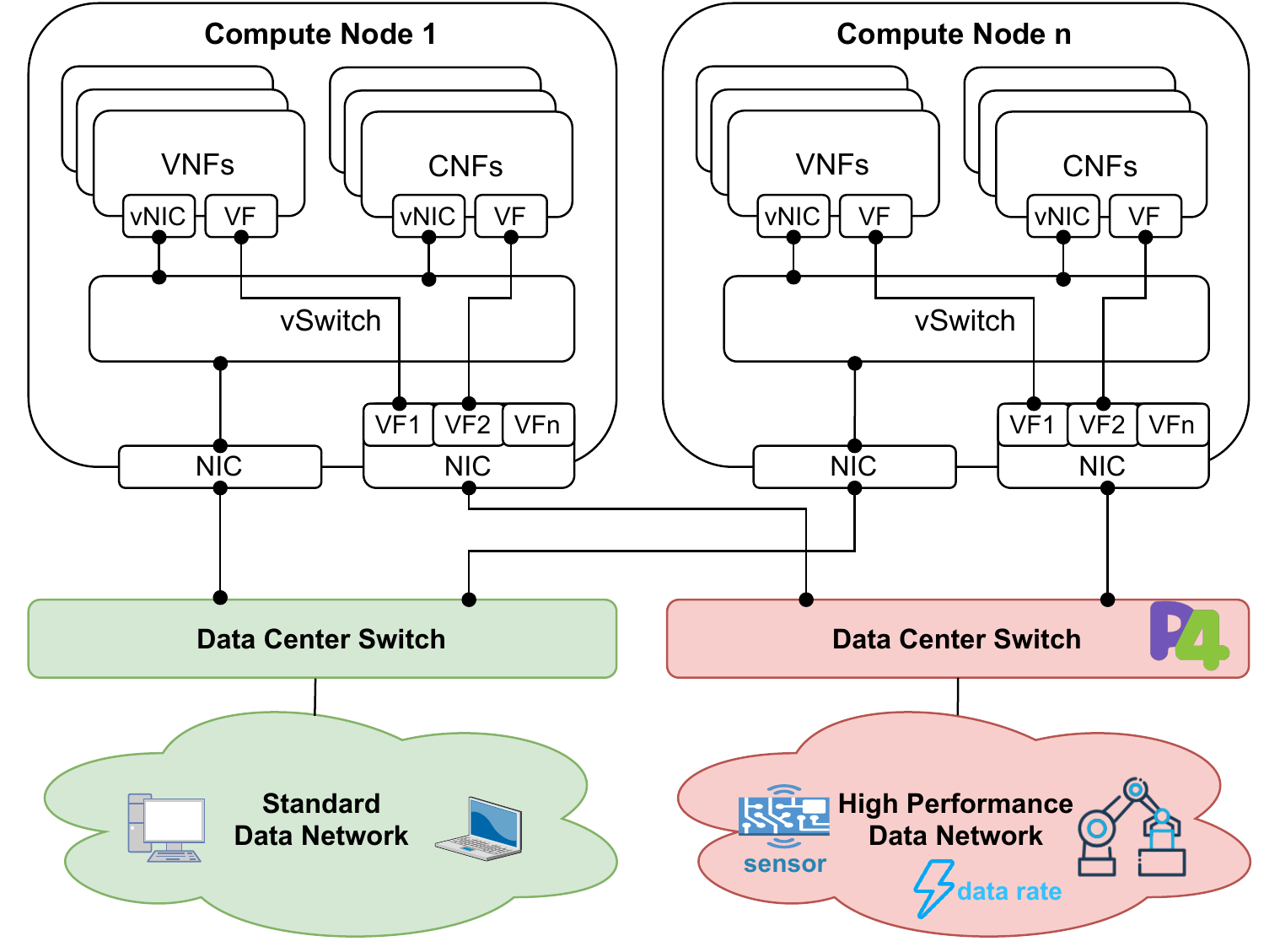}
}
\caption{Proposed system architecture to integrate P4NFs.}
\label{fig:arch}
\end{figure}

As part of the NFV architecture, OpenStack and Kubernetes are used as Virtualized Infrastructure Managers (VIMs) and Open Source MANO (OSM) is used as the Management and Orchestration (MANO) element. The NFVI is composed of several compute nodes that form the OpenStack cloud and the Kubernetes cluster, as well as of the switching infrastructure to interconnect those nodes, which includes both regular and P4 switches. OSM can deploy VNFs and CNFs on top of OpenStack and Kubernetes respectively, whereas P4NFs are employed to configure the P4 switches. These switches are registered in OSM as Physical Deployment Units. Connectivity between VNFs and CNFs can be provided through the regular top of the rack switches and/or through the P4 switches. To make this possible, the compute nodes have network interfaces connected to both types of switches, enabling the VNFs/CNFs to connect to any network. The VNF/CNF interfaces towards the P4 switches are SR-IOV Virtual Functions (VFs). To achieve isolation between tenants ---i.e. between NSs of different tenants---, separate OpenStack VLAN networks are assigned to each tenant. This proposed architecture is depicted in Fig. \ref{fig:arch}.

To design a P4NF, the user provides a reference to the desired functions in a P4NF metadata file, such as simple layer 2 forwarding, or layer 3 routing plus a custom layer 3 firewall. These functions are extracted from a shared repository of P4 code in which they are implemented in a modular way. A user may upload a new function that is not yet stored in the repository to use it in the P4NF. The requested functions are contrasted with the current state of the P4 pipeline, and the switch is reconfigured in order to upgrade to a new state that integrates the requirements of the new P4NF, which involves recompiling the DP and reconfiguring the CP. Every P4NF must include the parsing of Ethernet and VLAN to achieve the aforementioned tenant-level isolation.

\begin{figure}[tb]
\centerline{
\includegraphics[width=1\linewidth]{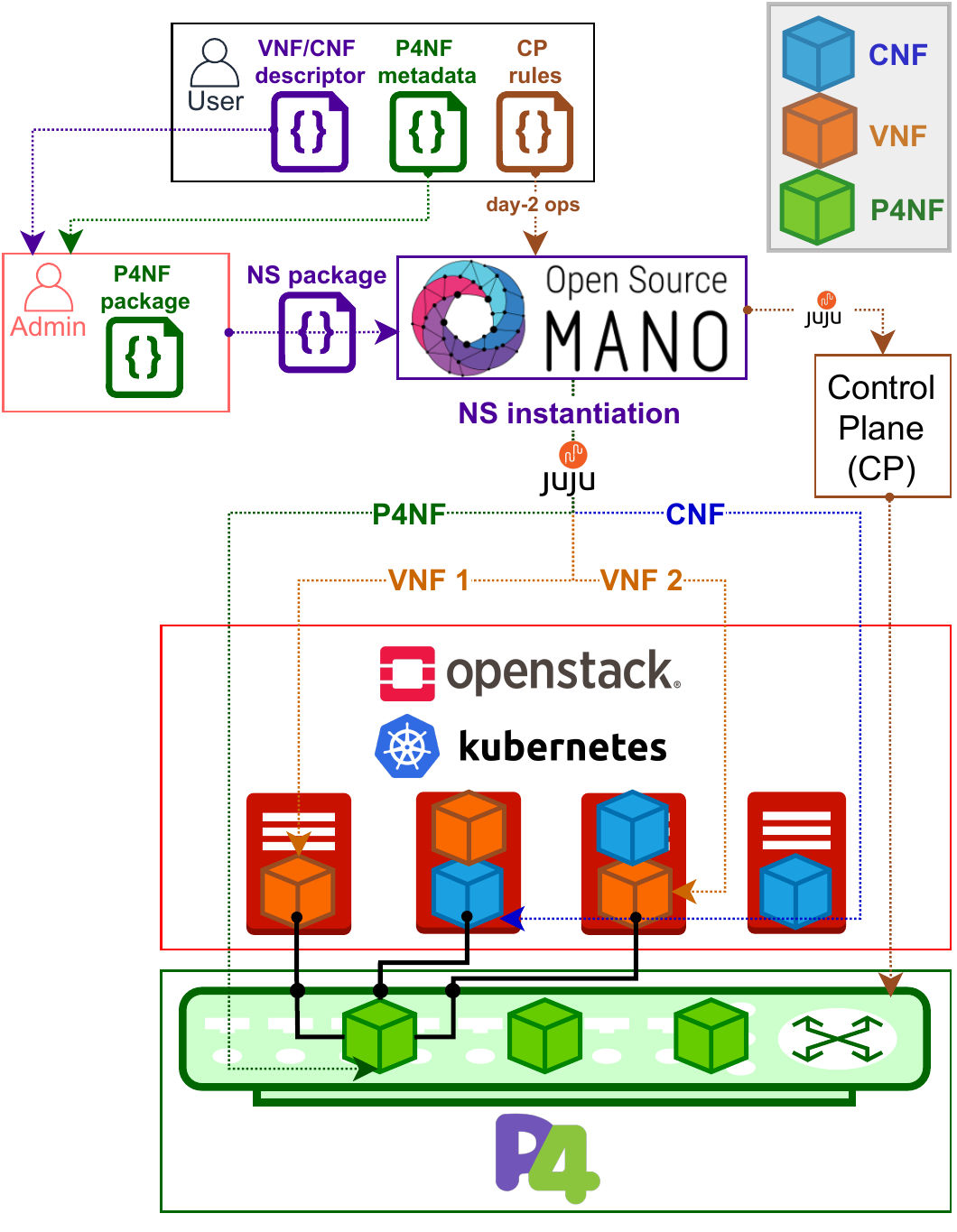}
}
\caption{P4-based NS onboarding and instantiation process.}
\label{fig:ns}
\end{figure}

The deployment of these P4NFs leverages OSM's support of Juju charms, which encapsulate a set of Python scripts that can be executed towards the target of the P4NF, i.e. a P4 switch. Apart from the P4NF, the user encapsulates the desired VNFs/CNFs into an NS. The P4NF designer provides the network administrator with the P4NF metadata file plus the VNFs/CNFs, so that the latter builds the complete package and instantiates it. Furthermore, Juju charms can be used to implement day-2 operations that allow the P4NF designer to directly modify the CP rules at any time once the NS is created. An example of the onboarding, instantiation, and lifecycle management processes of a composite NS is illustrated in Fig. \ref{fig:ns}.

\subsection{Low-level design}
This subsection presents a low-level description of the proposed deployment methodology for DP acceleration in NFV. Specifically, a workflow for the lifecycle management of the envisioned NSs is analyzed, detailing the procedures involved in each task. 

The framework hereby presented provides three main features that enable the description, instantiation, and management of the P4-enabled NSs. It provides \textbf{(a) multi-tenancy}, so the connectivity between NFs is restricted to NSs connected to the same OpenStack VLAN network, and, thus, to NSs belonging to the same tenant. It also offers \textbf{(b) DP composition} to describe P4 functions at NS level and then combine them to create a single P4 program that comprises the P4 functions of all running NSs ---even those from different tenants. In this way, the DP must be recomposed each time an NS is instantiated or terminated. The DP composing is implemented according to \cite{microp4,p4click}. Lastly, it supports \textbf{(c) DP slicing} to implement different and isolated packet processing pipelines for each NS that is configured when composing the DP.

A standard NS is formed by a combination of VNFs/CNFs while a P4-based NS includes a P4NF, which describes the configuration of the custom DP dedicated to that NS. Therefore, a new workflow must be defined to accommodate the management of the lifecycle of these P4NFs. In this case, Juju is in charge of the lifecycle management of the P4NFs, being responsible for their instantiation and termination. Thus, a P4NF is a PNF that points to the physical P4 switch and implements the required Juju charms for its configuration. 

To achieve these features, maintaining different types of state information is crucial. This leads to the following requirements:
\begin{itemize}
    \item Minimal Layer 2 (L2) forwarding requires the P4 switch to know the MAC addresses connected to each physical port. Traffic from different VNFs/CNFs could come into the same switch port as they could be hosted by the same compute node. A static match-action table (MAT), \textit{forward\_l2}, is used for this purpose. Each compute node SR-IOV VF is preconfigured with a static MAC address, and the entries with each VNF/CNF MAC address to switch port mapping are preloaded into the aforementioned table.
    \item Multi-tenancy and DP slicing support require maintaining a record in an external database of which DP functions are in use by each NS. This information is stored in the \textit{ns\_functions} table, which is consulted whenever the DP needs to be recomposed.
    \item Upon the instantiation or termination of a P4NF that requires recompiling the DP, all CP information is lost as the DP tables are destroyed. For this reason, an image of the current CP state must be kept up to date in an external database. The \textit{cp\_rules} table is introduced.
    \item The P4 code used by any P4NF needs to be stored in a public repository. This includes the code that would be required by a new P4NF, in order to pull said code during the instantiation of the P4NF. This repository, named \textit{P4 function repository}, allows any user to upload their custom DP functions so they are beforehand checked.
\end{itemize}

When an NS is instantiated, VNFs and CNFs are configured following the standard lifecycle operations defined in OSM. For the P4NF, the code defined in the Juju charms performs the following lifecycle operations to instantiate the P4NF:
\begin{enumerate}
    \item \textbf{Check required P4 functions}: The P4NF metadata file, which is part of the Juju charms, includes a list of the P4 functions that the DP must implement to support the P4NF.
    \item \textbf{Update \textit{ns\_functions} table}: As the P4 switch cannot run more than a single DP P4 program at once, this database table is required in order to obtain the complete list of the running P4 functions and recreate the program in the following step. Likewise, the table must be updated to reflect the DP pipeline state after the introduction of the new P4NF.
    \item \textbf{Compose P4 program}: This step consists in combining different P4 functions to create a single P4 program that configures the DP pipeline. Even if the new NS only requires P4 functions that already exist in the DP, a new DP needs to be composed, as the tables must be unique to restrict access to each DP table from the CP. This process is challenging and requires specific P4 coding rules to achieve composable P4 functions. The literature provides different approaches to accomplish this task \cite{microp4,p4click}. To use custom P4 functions, they must be uploaded to the \textit{P4 function repository}, which assures their "composability".
    \item \textbf{Compile P4 program}: The Juju charms establish an active connection to the P4 switch to load, compile, and run the P4 program. During this step, the P4 switch is not operational until the new program is running.
    \item \textbf{Configure switch ports}: in case the DP was recompiled, it is necessary to bring up all the ports that are in use by the switch. This is a fixed task, as all the ports with a physical connection to a compute node need to be enabled.
    \item \textbf{Update CP}: Finally, the Juju charms perform a basic CP configuration to provide L2 connectivity between the VNFs/CNFs specified in the NS. This is achieved by having the VNFs/CNFs send their MAC addresses to the P4NF through Juju. These parameters are then used as input parameters for the CP to configure the \textit{forward\_l2} MAT. On the other hand, the \textit{cp\_rules} database table needs to be consulted to restore the most recent state of the CP. 
\end{enumerate}

Additionally, CP operations that involve filling owned DP tables are allowed through the NF day-2 lifecycle operations supported by OSM and Juju. This includes both setting up the initial CP rules after the P4NF instantiation, and inserting new rules on demand. The exception to this is the \textit{forward\_l2} MAT, which, despite containing entries for all users, is managed by the network administrator only. Each day-2 operation is registered in the \textit{cp\_rules} database to keep it up to date.

The termination of the NS must be considered as part of the lifecycle management. Therefore, the code defined in the charms performs the following operations to remove the P4NF:
\begin{enumerate}
    \item \textbf{Update \textit{ns\_functions} table}: All active P4 functions are retrieved from the \textit{ns\_functions} table. Those functions that are no longer needed after the termination of the P4NF are removed from the table.
    \item \textbf{Compose P4 program}: Similarly to the NS instantiation, the P4 program is reconfigured. The no longer needed P4 functions after the termination of the P4NF are excluded to build the new P4 program. This includes the removal of unused MATs. This process is done to achieve the simplest possible DP pipeline.
    \item \textbf{Compile P4 program}: The same as in the NS instantiation workflow is applied.
    \item \textbf{Configure switch ports}: The same as in the NS instantiation workflow is applied.
    \item \textbf{Update CP}: The entries in the \textit{forward\_l2} MAT for the terminated P4NF are discarded. In addition, CP changes triggered through day-2 operations need to be reverted during the reconstruction of the CP. This is done by looking up the \textit{cp\_rules} table.
\end{enumerate}
\section{\uppercase{Conclusions and future work}}

DP acceleration in NFV is a promising idea, as the dynamic reconfiguration of the DP in an NFV-based environment grants network service providers with powerful possibilities. However, there are still some challenges that must be overcome in order to optimize the system for production environments. We hereby highlight and discuss the following three challenges.

\textbf{The first challenge} is sharing P4 MATs among tenants. This might seem a good option, as it results in better usage of resources and optimization of the P4 program. However, if a table is shared between two tenants, one could change the entries that correspond to the other tenant and illegitimately modify its behavior. In fact, unauthorized changes in the communication among VNFs could be a potential attack vector inside the system. The proposed solution has been to duplicate the tables for each tenant or NS, so that each of them only has access to its own tables and cannot modify other equivalent tables.

\textbf{The second challenge} is concerned with the duplication of parser code in P4 programs. The P4 program in the switch implements P4 functions in a modular way, only composing and compiling P4 functions required by instantiated NSs. However, multiple NSs may have partially concurrent needs, or, in other words, share one P4 function but require others that are different among them. This leads to different parsing requirements of those NSs. To put this in context, while two NSs might use the same static L2 forwarding table, one NS may just need to parse the Ethernet header, and the other one may need to parse even the application layer ---in order to execute other actions in a later phase. Therefore, a solution to this could be the duplication of parsers ---one for each NS---, or having a "super-parser" that is able to fulfill any parsing requirement of all instantiated NSs, using only the necessary parts for each case. 

\textbf{The third challenge} is related to the service downtime that arises due to the recompilation of the P4 program during the instantiation and deletion periods of NSs. Each time one of these processes is executed, the DP must be recomposed and recompiled, interrupting the service in the compilation part. This means that all deployed NSs lose connectivity in that period, which entails an availability issue. Approaches similar to \cite{microp4,dppx} could be studied to address this issue, in which the switch enters in a "slow-mode" state while the new DP is being recompiled, but does not stop serving with the current DP. Even though experiencing service downtime might not be reasonable for critical services, it could be for experimental environments that are not significantly affected by it. 

Therefore, our research shows that there are several methods under investigation that can provide solutions to the aforementioned challenges, and advance towards a fully operational system. There is still more work to do, mainly focusing on the challenges analyzed in this section, but we think that the overall approach is promising and may change the current service deployment system in NFV environments.
\section*{\uppercase{Acknowledgements}}

This work was supported in part by the Spanish Ministry of Science and Innovation through the national project (PID2019-108713RB-C54) titled "Towards zeRo toUch nEtwork and services for beyond 5G" (TRUE-5G), and in part by the "Smart Factories of the Future" (5G-Factories) (COLAB19/06) project.
\bibliographystyle{IEEEtran}

\IEEEpeerreviewmaketitle

\end{document}